\documentclass{aa}  

\usepackage{graphicx}
\usepackage{txfonts}
\begin{document} 

\title{Disproval of the validated planets K2-78b, K2-82b, and K2-92b}
\subtitle{The importance of independent confirmation of planetary candidates}

\author{
  J.~Cabrera\inst{\ref{DLR}}
  \and S.C.C.~Barros\inst{\ref{CAUP}}
  \and D.~Armstrong\inst{\ref{Warwick}}
  \and D.~Hidalgo\inst{\ref{IAC},\ref{ULL}}
  \and N.~C.~Santos\inst{\ref{CAUP},\ref{UPorto}}
  \and J.~M.~Almenara\inst{\ref{Geneve},\ref{UGrenoble}}
  \and R.~Alonso\inst{\ref{IAC},\ref{ULL}}
  \and M.~Deleuil\inst{\ref{LAM}}
  \and O.~Demangeon\inst{\ref{CAUP}}
  \and R.~F.~D{\'i}az\inst{\ref{Universidad Buenos Aires},\ref{CONICET},\ref{Geneve}}
  \and M.~Lendl\inst{\ref{Graz}}
  \and J.~Pfaff\inst{\ref{TUB}}
  \and H.~Rauer\inst{\ref{DLR},\ref{TUB}}
  \and A.~Santerne\inst{\ref{LAM}}
  \and L.~M.~Serrano\inst{\ref{CAUP},\ref{UPorto}}
  \and S.~Zucker\inst{\ref{Tel-Aviv}}
}

\institute{
  Deutsches Zentrum f{\"u}r Luft- und Raumfahrt, Rutherfordstr. 2, D-12489 Berlin, Germany\\\email{juan.cabrera@dlr.de}\label{DLR}
  \and Instituto de Astrof{\'i}sica e Ci{\^e}ncias do Espa\c{c}o, Universidade do Porto, CAUP, Rua das Estrelas, 4150-762 Porto, Portugal\label{CAUP}
  \and University of Warwick, Department of Physics, Gibbet Hill Road, Coventry, CV4 7AL, UK\label{Warwick}
  \and Instituto de Astrof{\'i}sica de Canarias, V{\'i}a L{\'a}ctea s/n, E-38205 La Laguna, Tenerife, Spain\label{IAC}
  \and Departamento de Astro{\'i}sica, Universidad de La Laguna, E-38206 La Laguna, Tenerife, Spain\label{ULL}
  \and Departamento de F{\'i}sica e Astronomia, Faculdade de Ci{\^e}ncias, Universidade do Porto, Rua do Campo Alegre, 4169-007 Porto, Portugal\label{UPorto}
  \and Observatoire de Gen{\`e}ve, Universit{\'e} de Gen{\`e}ve, 51 chemin des Maillettes, 1290 Versoix, Switzerland\label{Geneve}
  \and Universit{\'e} Grenoble Alpes, CNRS, IPAG, F-38000 Grenoble, France\label{UGrenoble}
  \and Aix Marseille Univ., CNRS, LAM, Laboratoire d’Astrophysique de Marseille, Marseille, France\label{LAM}
  \and Universidad de Buenos Aires, Facultad de Ciencias Exactas y Naturales. Buenos Aires, Argentina\label{Universidad Buenos Aires}
  \and CONICET - Universidad de Buenos Aires. Instituto de Astronom\'ia y F\'isica del Espacio (IAFE). Buenos Aires, Argentina\label{CONICET}
  \and Space Research Institute, Austrian Academy of Sciences, Schmiedlstr. 6, A-8042 Graz, Austria\label{Graz}
  \and Department of Astronomy and Astrophysics, Berlin University of Technology, Berlin, Germany\label{TUB}
  \and Department of Geosciences, Raymond and Beverly Sackler Faculty of Exact Sciences, Tel-Aviv University, Tel Aviv 6997801, Israel\label{Tel-Aviv}
}

   \date{Received ; accepted}

 
   \abstract
   {Transiting super-Earths orbiting bright stars in short orbital
     periods are interesting targets for the study of planetary
     atmospheres.} 
   {While selecting super-Earths suitable for further characterization
     from the ground among a list of confirmed and validated
     exoplanets detected by K2, we found some suspicious cases
     that led to us re-assessing the nature of the detected transiting
     signal.} 
   {We did a photometric analysis of the K2 light curves and 
     centroid motions of the photometric barycenters.}
   {Our study shows that the \emph{validated} planets K2-78b, K2-82b,
     and K2-92b are actually not planets but background eclipsing
     binaries. The eclipsing binaries are inside the Kepler
     photometric aperture, but outside the ground-based high
     resolution images used for validation.} 
   {We advise extreme care on the validation of candidate planets
     discovered by space missions. It is important that all the
     assumptions in the validation process are carefully checked. An
     independent confirmation is mandatory in order to avoid wasting
     valuable resources on further characterization of non-existent
     targets.} 

   \keywords{methods: data analysis -- techniques: photometric -- eclipses -- planets and satellites: detection -- planets and satellites: individual: K2-78b, K2-82b, K2-92b}

   \maketitle
%
\section{Introduction}

The largest fraction of the 3\,580 transiting planets known to
date \citep[i.e.][\url{http://exoplanet.eu/}]{schneider2011} have
been found by space missions like CoRoT~\citep{baglin2006} and
especially by Kepler~\citep{borucki2010a} and K2~\citep{howell2014}.  
However, only a small fraction of these planets have been
independently confirmed with radial velocity (RV) measurements.
Fortunately, the extraordinary photometric precision of space-borne
observatories has allowed a validation process of planetary candidates
based on statistical studies of the distribution of planetary
populations and the most common false positive
scenarios~\citep{torres2011,morton2012,diaz2014a,santerne2015}, rather
than on an independent characterization of the planetary properties
with spectroscopic measurements.

The photometric analysis of the light curve to confirm the planetary
nature of a transiting candidate is a standard step of the ranking
process of planetary candidates~\citep{armstrong2017a}.
The simplest steps include the search for secondary eclipses or
ellipsoidal variations (also referred as out-of-transit variation)
revealing the stellar nature of the transiting body.
The analysis of the chromatic light curves in
CoRoT~\citep{almenara2009} or the centroid motion analysis in
Kepler~\citep{batalha2010b} are also powerful tools to reject
contaminating eclipsing binary scenarios. 
However, these steps are primarily used as a tool to veto candidates
before any time consuming photometric or spectroscopic follow-up
observations are carried out.

With Kepler, the validation of candidates which are too faint to be
observed with ground-based observatories, or whose expected mass was
estimated too low to be detectable with current instruments, gave a
step forward.
More sophisticated analysis tools like BLENDER~\citep{torres2011} or
PASTIS~\citep{diaz2014a,santerne2015} succeeded in 
rejecting all possible non-planetary scenarios compatible with the
properties of the planetary candidate found in the light curve.
These tools could make efficient use of all available information
(stellar properties, Galaxy models, complementary observations in
different wavelengths, etc.) to secure the posterior of the
hypothesis that the candidate was indeed a planetary companion.
Needless to say, the performance of these tools is as good as the
reliability of the information used in the analysis of the
hypothesis.

Recently,~\citet{crossfield2016} used the validation tool
VESPA~\citep{morton2015vespa,morton2016} to confirm the planetary
nature of 104 planets observed by K2. 
In particular, they validated the planetary nature of K2-78b
(EPIC~210400751), K2-82b (EPIC~210483889), and K2-92b
(EPIC~211152484), all with a false-positive probability of less than 
1\%.
We were interested in the study of these targets from an observational
point of view.
They are super-Earths receiving large amount of stellar
irradiation, having high equilibrium temperatures and consequently
relatively large scale heights, orbiting relatively bright stars,
favorable for further characterization.
Unfortunately, in this paper we show that these validated
super-Earth-sized planets are actually blended eclipsing
binaries.
This is not the result of a statistical fluctuation, but the
consequence of not including all the available information about these
targets, resulting in a wrong evaluation of the false-positive
probability.

\section{The falsified planets}

\citet{crossfield2016} published a study where they presented 197
candidates found in the K2 data together with an ambitious
ground-based follow-up programme, including photometric analysis, high
angular resolution imaging, and stellar spectroscopy which lead
them to validate 104 planets, i.e. statistically confirm their
planetary nature, 64 of them validated for the first time.

Our study shows that 3 of these new 64 validated planets, all with
false positive probabilities less than 1\% as estimated
by~\citet{crossfield2016}, are actually blended eclipsing 
binaries. 

\subsection{K2-92b -- EPIC 211152484}

Many of the new candidates validated by~\citet{crossfield2016} are
small planets (below 2 Earth radii) in close orbits around relatively
bright stars, which makes them interesting targets for atmospheric
characterization.
One of the most interesting targets for our team was K2-92b 
(EPIC~211152484), which drove us to a closer examination of its
properties prior to further theoretical modelling and
characterization with ground-based facilities.

K2-92b was validated by~\citet{crossfield2016} as a planet with an
orbital period of 0.7018180 days, a radius of 2.56 Earth radii and a
false positive probability of less than 0.12\% orbiting a star of
magnitude 12.136 in the Kepler pass-band.
During our study, we compared the transit depth as a function of the
size of the photometric aperture using data reduced with the pipeline
by~\citet{vanderburg2014}.
We found out that the transit depth depended strongly on the size of
the aperture used to extract the photometry.  

\begin{figure}
  \centering
  \includegraphics[%
  width=0.9\linewidth,%
  height=0.5\textheight,%
  keepaspectratio]{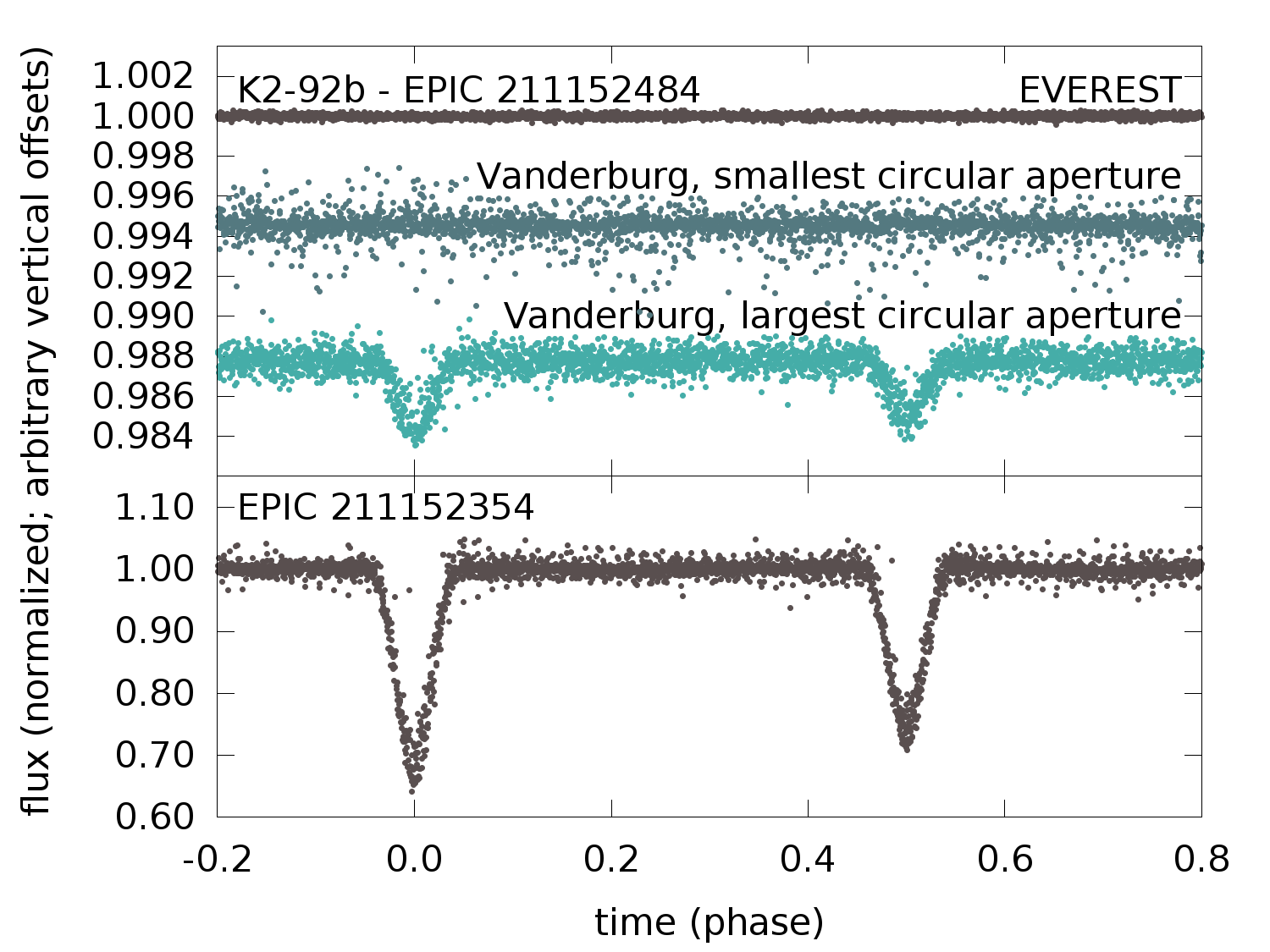}
  \caption{Folded light curve of K2-92b with different apertures (top)
    and of EPIC~211152354 (bottom) folded at 1.4 days orbital
    period. See text for details.}
  \label{K2-92b_folded}
\end{figure}

If there is a neighbouring star close to the target, one would expect
the transit depth to decrease when enlarging the aperture, due to the
inclusion of background light or contaminating light from the
neighbouring star.
However, in the case of K2-92b we observed the opposite effect. 
The largest transit depth corresponded to the largest aperture, which
is a clear sign that the real transit signal comes actually from the
background source.
We compare in the top part of Fig.~\ref{K2-92b_folded} the photometry
of K2-92b extracted with Everest~\citep{luger2016,luger2017} and with
the code by~\citet{vanderburg2014} folded at twice the orbital period
quoted by~\citet{crossfield2016}. 
The Everest data do not show any transit feature, same as the
Vanderburg code with the smallest aperture.
However, the largest aperture from Vanderburg does show the expected
signal at the right period, only with a larger depth (about 0.4\%
compared to the tabulated 0.03\%).

We folded the data at twice the orbital period quoted in the
validation paper because we considered that the transit depth
differences between odd and even transit events at 0.7 days period are
significant.
The analysis shows that the star responsible for the signal is an
eclipsing binary with different depths for the primary and secondary
eclipses, at about 1.4 days orbital period. 
In this particular case, the star responsible for the variability
observed in the K2 light curve is a faint (G band 17.045,
\citealt{brown2016GaiaDR1}) star (with identification EPIC~211152354)
about 15 arcseconds south east of the main K2 target (see bottom part
of Fig.~\ref{K2-92b_folded}) showing eclipses of 35\% depth. 

The analysis of the centroid motion has been proposed as an useful
tool to reject false positive scenarios~\citep{batalha2010b}.
Although in this case the source of the contaminant has been clearly
identified, we decided to use the pipeline POLAR that is based on the
CoRoT imagette pipeline to calculate the centroid motion of K2-92 in
phase with the transit signal. 
A full description of the POLAR pipeline was presented in
\citet{barros2016}. 
Briefly, the centre of light is calculated using the Modified Moment
Method by~\citet{Stone1989} then the line of sight of
the Kepler satellite is subtracted to obtain the centroid motion of each star. 
This pipeline has been used to discover and characterize several K2
exoplanet discoveries
e.g. \citet{barros2015}.  
The reduced light curves up to campaign 6 are publicly available
through the MAST (\url{https://archive.stsci.edu/prepds/polar/}).

\begin{figure}
  \centering
  \includegraphics[%
  width=0.9\linewidth,%
  height=0.5\textheight,%
  keepaspectratio]{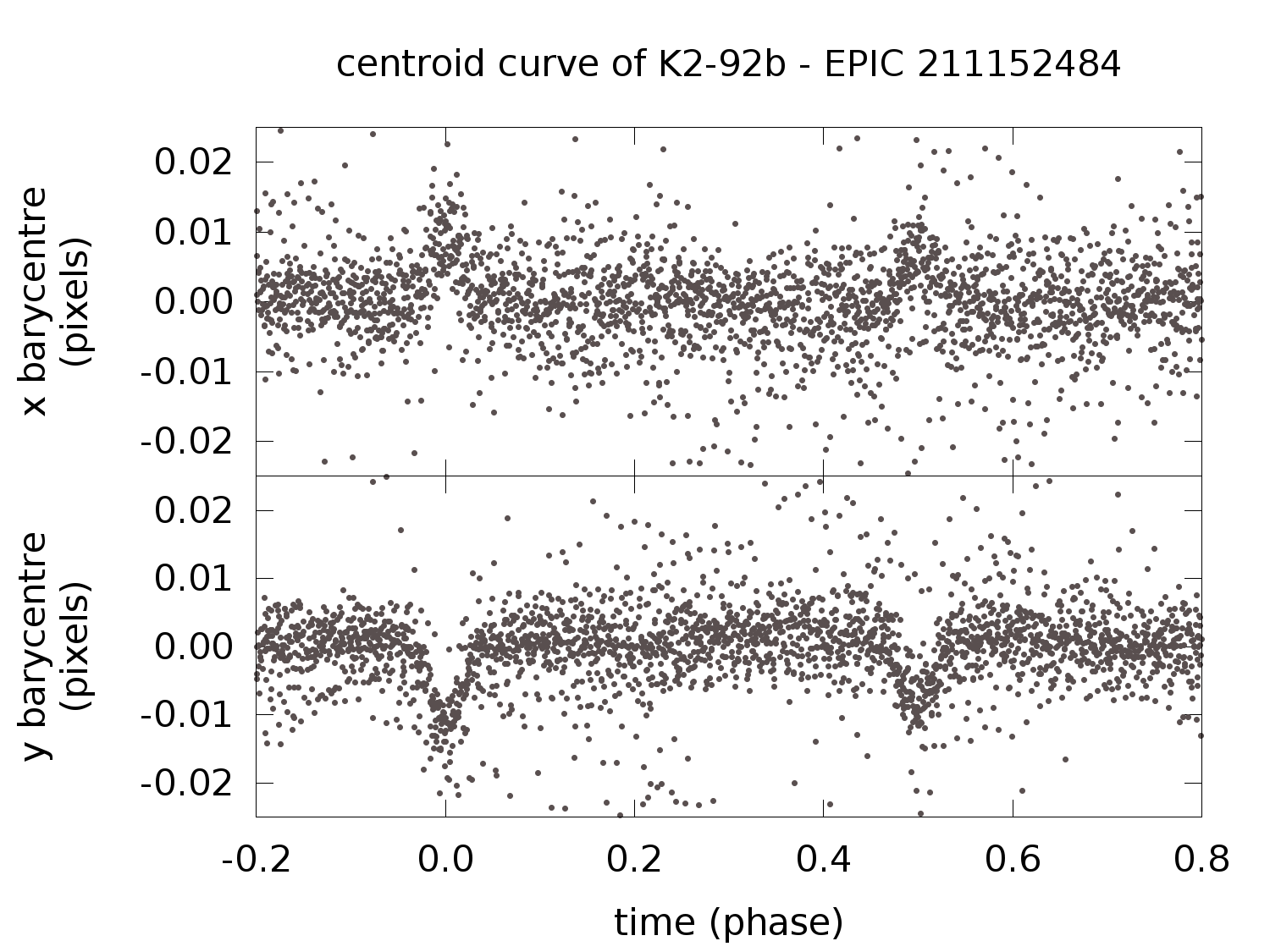}
  \caption{Time series of the centroid motion of K2-92 folded at the
    period of the photometric transit. See text for details.}
  \label{K2-92b_centroid}
\end{figure}

In Fig.~\ref{K2-92b_centroid} we show the centroid motion of K2-92b
for the x and y directions, phase folded on the 1.4 day orbital period
of the binary. It is clear that there exists a strong correlation
between the centroid motion and the transit phase, which we indicates
that a neighbouring star is the source of the signal.

We note that~\citet{adams2016a} also reported an unusual behaviour of
the transit depths of K2-92b. 
They mentioned stellar variability, debris clouds, or even a comet as
possible explanations for the irregular behaviour of the candidate. 
However, they failed to identify the eclipsing binary as the source of the
signal.

\subsection{K2-78b -- EPIC 210400751}

\begin{figure}
  \centering
  \includegraphics[%
  width=0.9\linewidth,%
  height=0.5\textheight,%
  keepaspectratio]{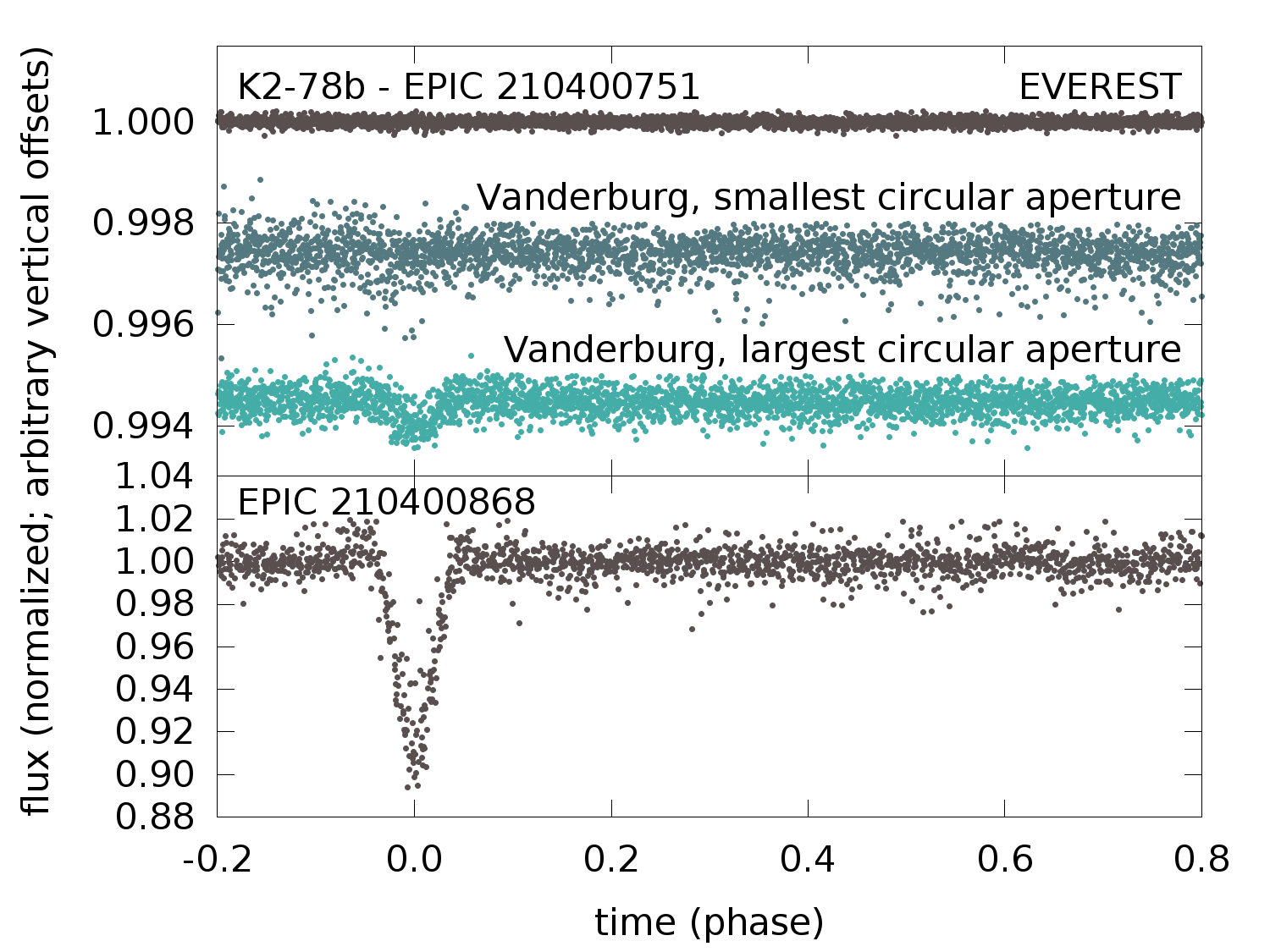}
  \caption{Folded light curve of K2-78b with different apertures (top)
    and of EPIC~210400868 (bottom) folded at 2.3 days orbital
    period. See text for details.}
  \label{K2-78b_folded}
\end{figure}

K2-78b was validated by~\citet{crossfield2016} as a planet with an
orbital period of 2.29016 days, a radius of 1.42 Earth radii and a
false positive probability of less than 0.31\% orbiting a star of
magnitude 11.892 in the Kepler pass-band.
We proceeded in the same way as for K2-92b (see
Fig.~\ref{K2-78b_folded}) and show that the star responsible for the
variability (with eclipses of 10\% depth) lies to the north of the
main target and is about 4 magnitudes fainter (EPIC~210400868). 

\subsection{K2-82b -- EPIC 210483889}

\begin{figure}
  \centering
  \includegraphics[%
  width=0.9\linewidth,%
  height=0.5\textheight,%
  keepaspectratio]{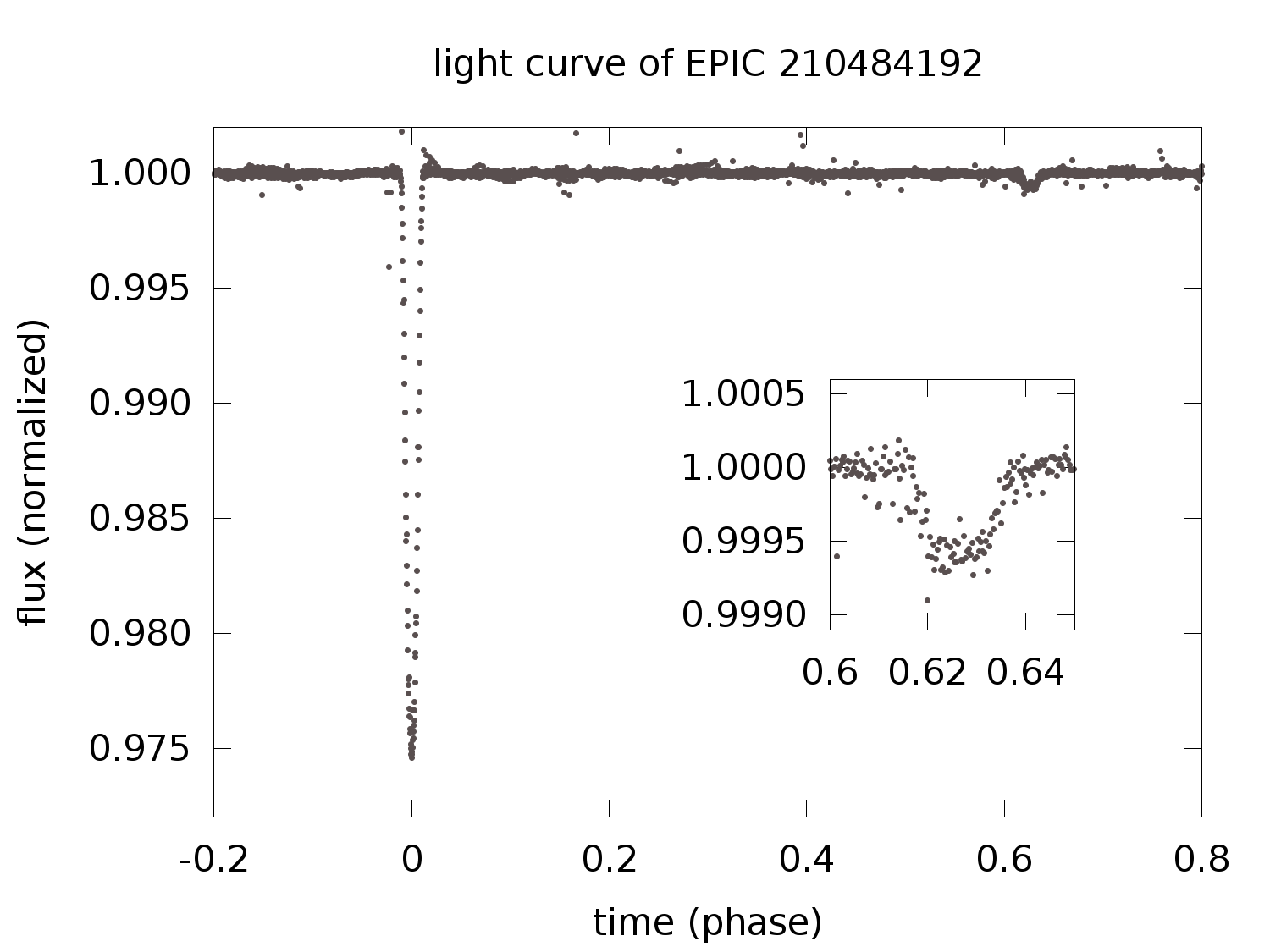}
  \caption{Folded light curve of EPIC~210484192 at the ephemeris
    published by~\citet{crossfield2016} for K2-82b. The close-up shows
    the phase around the secondary eclipse. See text for details.} 
  \label{epic210484192_folded}
\end{figure}

K2-82b was validated by~\citet{crossfield2016} as a planet with an
orbital period of 7.195834 days, a radius of 2.6 Earth radii and a
false positive probability of less than 0.059\% orbiting an M dwarf of
magnitude 13.519 in the Kepler pass-band. 
The transit depth reported by~\citet{crossfield2016} is about 2.0\%,
but the fact that the EPIC target is an M dwarf (0.17$R_\mathrm{Sun}$)
results in a very small planetary radius (2.6$R_\mathrm{Earth}$).
In this case, our analysis of the Everest light curve shows a primary
eclipse of 2.5\% depth and a clear secondary eclipse at phase 0.62
(the eclipsing binary being eccentric) in the light curve, which is
incompatible with the occultation of a planetary object (see
Fig.~\ref{epic210484192_folded}).
It is unclear why the signal of the secondary eclipse was ignored in
the validation process. 
The source of the signal is not the M dwarf, but a bright star (V=9.0)
to the north of the main target (EPIC~210484192), which got its own
aperture in the C4 campaing of K2~\citep{armstrong2016a}.



\section{Discussion}

Our result shows that, though planet validation techniques are useful
tools, great care needs to be taken to correctly validate candidate
planets discovered by space missions.
\citet{crossfield2016} made a sound statistical study and a careful
and detailed ground-based characterization of the targets, including
high angular resolution imaging, but they failed to look for possible
contaminants a few arcseconds away from the targets.
In the cases mentioned above, the contaminants were too far away to be
included in the field of view of the high resolution image and they
were not considered further in the analysis. 

The reliability of a statistical study is only as good as the
understanding of the contamination sources.
Here we show i) that validation methods applied to these targets
by~\citet{crossfield2016} underestimate the impact of background
contaminants and consequently, ii) the planet likelihood estimates are
not representative of the true nature of the candidates in these cases.
{\bf We insist that this is not the result of a failure of the design
  of the validation procedure, but the result of an incorrect
  assessment of the impact on the photometry of neighbouring sources.
  Our results can be used to improve the performance of planet
  validation techniques.}

Checking the light curves using different aperture sizes is a common
validation step made in ground-based transit surveys. 
In this paper we show that it can also reveal false positive scenarios
in space-borne surveys, saving valuable follow-up resources.
We suggest to introduce these tests in the pipelines of
TESS~\citep{ricker2015} and PLATO~\citep{rauer2014}.

The use of validated planets might be justified for statistical
studies of large populations, as long as the theoretical studies can
deal with a certain contamination which might not be completely 
described by the false-positive values of individual systems.
The reliable statistical validation of individual systems is complex
and costly, and one could risk saying that the detailed study of
individual planetary systems requires the use of independently
confirmed planets with RV measurements or, as a minimum, significant
independent evidence, like additional planetary companions in the
system or transit-timing variations consistent with the planetary
scenario~\citep{barros2013}.
The risk is wasting telescope time and modelling efforts in false
positive scenarios.
Furthermore, if a significant number of particularly valuable "false
positive" planet candidates are not discarded by validation
procedures, their inclusion in statistical analysis studies of planet
populations may be biased.

\begin{acknowledgements}
This publication was written by the International Team led by
J.C. on ‘Researching the Diversity of Planetary Systems’ at ISSI    
(International Space Science Institute) in Bern.
We acknowledge the financial support of ISSI and thank them for their
hospitality. 
SCCB, NCS, OD, and LS acknowledge support by Funda\c{c}\~ao para a
Ci\^encia e a Tecnologia (FCT, Portugal) through the research grant
through national funds and by FEDER through COMPETE2020 by grants
UID/FIS/04434/2013 \& POCI-01-0145-FEDER-007672 and
PTDC/FIS-AST/1526/2014 \& POCI-01-0145-FEDER-016886. 
S.B. and N.C.S. also acknowledge support from FCT through Investigador FCT 
contracts nr. IF/01312/2014/CP1215/CT0004 and
IF/00169/2012/CP0150/CT0002.  
DJA is funded under STFC consolidated grant reference ST/P000495/1.
DH and RA acknowledge the Spanish Ministry of Economy and
Competitiveness (MINECO) for the financial support under the Ram{\'o}n
y Cajal program RYC-2010-06519, and the program RETOS
ESP2017-57495-C2-1-R.
This paper includes data collected by the Kepler mission. Funding for
the Kepler mission is provided by the NASA Science Mission
directorate.
This work has made use of data from the European Space Agency (ESA)
mission {\it Gaia} (\url{http://www.cosmos.esa.int/gaia}), processed by
the {\it Gaia} Data Processing and Analysis Consortium (DPAC,
\url{http://www.cosmos.esa.int/web/gaia/dpac/consortium}). Funding
for the DPAC has been provided by national institutions, in particular
the institutions participating in the {\it Gaia} Multilateral Agreement.
\end{acknowledgements}

%
%

\bibliographystyle{bibtex/aa.bst}
\bibliography{bibl}

\end{document}